\documentclass[fleqn, 12pt] {article}
\usepackage{epsf}
\begin{document}
\title{Effect of magnetic field on temporal development of Rayleigh -Taylor instability induced interfacial nonlinear structure}
\author{M. R. Gupta\thanks{e-mail: mrgupta$_{-}$cps@yahoo.co.in}, Labakanta Mandal\thanks{e-mail: laba.kanta@yahoo.com}, Sourav Roy \thanks{e-mail: phy.sou82@gmail.com},
Manoranjan Khan\thanks{e-mail: mkhan$_{-}$ju@yahoo.com} \\
Deptt. of Instrumentation Science \& Centre for Plasma Studies
\\Jadavpur University, Kolkata-700032, India\\}
\date{}
\maketitle

\begin{abstract}
The effect of magnetic field on the nonlinear growth rate of
Rayleigh - Taylor instability induced two fluid interfacial
structures has been investigated. The magnetic field is assumed to
be parallel to the plane of the two fluid interface and acts in a
direction perpendicular to the wave vector. If magnetic field is
restricted only to either side of the interface the growth rate
may be depressed (may almost disappear) or be enhanced depending
on whether the magnetic pressure on the interface opposes the
instability driving pressure difference $g(\rho_h-\rho_l)$y or
acts in the same direction. If magnetic field is present on both
sides of the two fluid interface, stabilization may also take
place in the sense that the surface of separation undulates
periodically when the force due to magnetic pressure on two sides
are such as to act in opposite direction. This result differs from
the classical linear theory result which predicts that the
magnetic field parallel to the surface has no influence on the
growth rate when the wave vector is perpendicular to its
direction.
\end{abstract}
\newpage

\section*{I. INTRODUCTION}
Temporal development of nonlinear structures at the two fluid
interface consequent to Rayleigh -Taylor (RT) or Richtmyer -
Meshkov (RM) instability is of much current interest both from
theoretical and experimental point of view. The structure is
called a bubble if the lighter fluid pushes across the unperturbed
interface into the heavier fluid and a spike if the opposite takes
place. The importance of such instabilities arises in connection
with a wide range of problems ranging from astrophysical phenomena
such as Supernova remnant to Inertial Confinement Fusion (ICF). A
core collapse Super Nova (SN) is driven by an externally powerful
shock, and strong shocks are the breeding ground of hydrodynamic
instabilities like RT and RM instabilities. During the shock
transit phase, the RM instability is activated at each
discontinuity in the density profile of the star at the O-He and
He-H interface. After shock transit, hydrodynamic mixing continues
due to RT instability, as the denser layers are decelerated by
lower density outer layer.

In an ICF situation, ablation front of an imploding capsule is
subject to the RT instability because dense core is compressed and
accelerated by low density ablating plasma. RT instability
enhances the perturbation initiated by laser induced target non
uniformity and consequently the performance of ICF implosion may
be seriously affected. The dynamics of the instability of the
interface of two constant density non-conducting fluids and the
associated nonlinear structure has been studied by several authors
$^{\cite{jh94} -\cite{ss03}}$ using an expression near the tip of
the bubble or the spike up to second order in the transverse
coordinate following Layzer's approach $^{\cite{dl55}}$. The
fluids may also be ionized as in the astrophysical situation or
may get ionized through laser irradiation in laboratory condition.
Magnetic field generated by ponderomotive force can exist
$^{\cite{mk92}-\cite{rm98}}$ in such conducting (ionized) fluids
and have important influence on the growth or suppression of the
instabilities.

When the explosion of a Ia type supernova (SNIa) starts in a white
dwarf as a laminar deflagration at the center of the star, RT
instability begins to act $^{\cite{cr01},\cite{eg93}}$. The
burning velocity at these regimes can be described by fractal
model of combustion. In white dwarf, magnetic field with strength
upto $10^8\sim10^9$ G exist at the surface and the field near the
center may be $\sim$ 10 times greater. Rayleigh - Taylor
instability arising during type Ia supernova explosion is
associated with strong magnetic field. Since the magnetic field is
dipolar type the fluid propagates parallel to the field lines
(i.e., approximately along the direction of gravity) near the
magnetic pole while the field lines are transverse to the
direction of gravity at the magnetic equatorial region. Thus
magnetic field effect on RT instability may have important roles
to play whether the field lines are normal or parallel to the two
fluid interface (i.e., along or perpendicular to the direction of
gravity).

The effect of magnetic field on Rayleigh - Taylor instability has
been studied in detail previously by Chandrasekhar
$^{\cite{sc68}}$. When the magnetic field is normal to the surface
of separation of the two fluids, the RT instability is almost
unaffected by the magnetic field when the wave number 'k' of the
perturbation is small; but contrary to the purely hydrodynamic
case the growth rate does not increase indefinitely with 'k' but
tends to a saturation value as $k \to \infty$. Magnetic field
parallel to the direction of impulsively generated acceleration
$^{\cite{vw05}}$ is also shown to induce RM instability. This
however happens for sufficiently intense magnetic field and also
tends asymptotically to a saturation value.

In case the magnetic field is parallel to the surface of
separation it is found that according to linear theory there
exists no effect of the magnetic field on the instability
$^{\cite{sc68}}$  if the latter is perpendicular to the wave
vector $\vec{k}$. Non vanishing effect of transverse magnetic
field with $\vec{k}$ perpendicular to the zeroeth order magnetic
field is however found to exist in linear theory when
compressibility effect is taken into account $^{\cite{ls08}}$. The
growth rate is found to be lowered both for continuously
accelerated (RTI) and impulsively accelerated (RMI) two fluid
interface $^{\cite{rs03} -\cite{mg09}}$ when $\vec{k}$ has
component parallel to the magnetic field. The nature of the
depression has close resemblance to that due to surface tension
$^{\cite{sc68}}$ and also has useful application in astrophysical
context $^{\cite{cr01} ,\cite{bi95}}$.

 The present paper is addressed to the problem of the time
 development of the nonlinear interfacial structure caused by
 Rayleigh Taylor instability in presence of a magnetic field
 parallel to the surface of separation of the two fluids . The
 wave vector is assumed to lie in the same plane and perpendicular
 to the magnetic field. With such a geometry there is no effect of
 the magnetic field in the classical $^{\cite{sc68}}$ linear approximation. However, it is no longer the case when
 linearization restriction is lifted. This may be understood from the following
 consideration.

 In presence of magnetic field, there exists the magnetic pressure
 in addition to the usual hydrodynamic pressure. As a result the RT
 instability driving pressure difference $g(\rho_h-\rho_l)$y is
 changed by the inclusion of the magnetic pressure difference
 $(1/2\mu)(B_h^2-B_l^2)$ [the suffix h(l) correspond to the dynamical variable associated with the heavier (lighter)
 fluid]. This has the consequence that the growth rate may be
 enhanced or depressed according as the extra contribution is
 either positive or negative . Moreover, as we shall see there may also occur
 stabilization in the sense that the surface of separation executes periodic undulation resulting from time lag in the
 temporal variation of $B_h$ and $B_l$. It is interesting to note
 that these are entirely nonlinear effects and disappear in the
 linear approximation.

    Section II deals with the basic MHD equation together
with the geometry involved. The fluid is assumed inviscid and
perfectly conducting and the fluid motion to be one of potential
type motion. The investigation of the nonlinear aspect of the
mushroom structure of the two fluid interface is facilitated by
Bernoulli's equation - the first integral of the equation of
motion of the magnetofluid obtained with the help of the magnetic
field geometry. The kinematical and dynamical boundary conditions
holding at the two fluid interface are set forth in section III.
The set of equations describing the temporal development of the RT
instability induced nonlinear structures at the interface are
derived in section IV. As these equation are not amenable to
solution in closed analytic form, the results are obtained by
numerical methods followed by graphical results and are presented
in section V. A summary of the results is given in section VI.

\section*{II. BASIC EQUATIONS}
Assume that the undisturbed surface is $y=0$, the transverse
coordinates being represented by $x,z$. The heavier fluid (density
$\rho_h =$constant) occupies the region $y>0$ while the lighter
fluid (density $\rho_l =$constant) is in the region $y<0$; gravity
is taken to point along negative $y$ axis.

 As shown in Fig. 1, the magnetic field is taken along the z
 direction:
\begin{eqnarray} \label{eq:1}
\vec{B}=\hat{z}B_h(x,y,t); \qquad \qquad y>0 \\ \nonumber \quad
=\hat{z}B_l(x,y,t); \qquad \qquad y<0\end{eqnarray}
\begin{eqnarray} \label{eq:2}
\mbox {so that }  \vec\nabla.\vec{B}=0
\end{eqnarray}
automatically.
   The mushroom shaped perturbation interface which is
called a bubble or a spike according as the lighter fluid pushes
into the heavier fluid or the opposite is taken to have a
parabolic form :
\begin{eqnarray} \label{eq:3}
y(x,t)=\eta_0(t)+\eta_2(t)x^2
\end{eqnarray}
Thus we have
\begin{eqnarray}\label{eq:4}
\mbox{for a bubble:}\qquad\qquad \eta_0>0 \quad \mbox{and} \quad
\eta_2<0
\end{eqnarray}
\begin{eqnarray}\label{eq:5}
\mbox{for a spike:} \qquad\qquad \eta_0<0 \quad \mbox{and} \quad
\eta_2>0
\end{eqnarray}

    For uniform density fluid the equation of continuity
$\mbox{} {\vec{\nabla.\vec{v}}}=0$ is satisfied for irrotational
fluid motion. Following Goncharov $^{\cite{vg02}}$ the velocity
potentials describing the irrotational motion for the heavier and
lighter fluids are assumed to be given by

\begin{eqnarray}\label{eq:6}
\phi_h(x,y,t)=a_1(t)\cos{(kx)}e^{-k(y-\eta_0(t))}; \quad y>0
\end{eqnarray}
\begin{eqnarray}\label{eq:7}
\phi_l(x,y,t)=b_0(t)y+b_1(t)\cos{(kx)}e^{k(y-\eta_0(t))}; \quad
y<0
\end{eqnarray}
\begin{eqnarray}\label{eq:8}
\mbox{with }\vec{v}_{h(l)}=-\vec{\nabla}\phi_{h(l)} .
\end{eqnarray}

The fluid motion is governed by the ideal magneto hydrodynamic
equations
\begin{eqnarray}\label{eq:9}
\rho_{h(l)}\left[\frac{\partial \vec{v}_{h(l)}}{\partial
t}+(\vec{v}_{h(l)}.\vec{\nabla})\vec{v}_{h(l)}\right]=-\vec{\nabla}p_{h(l)}
-\rho_{h(l)}\vec{g}+\frac{1}{\mu_{h(l)}}(\vec{\nabla}\times\vec
{B}_{h(l)})\times\vec{B}_{h(l)}
\end{eqnarray}
\begin{eqnarray}\label{eq:10}
\frac{\partial \vec{B}_{h(l)}}{\partial
t}=\vec{\nabla}\times[\vec{v}_{h(l)}\times\vec{B}_{h(l)}]
\end{eqnarray}
For magnetic field of the form given by Eq. (1)
\begin{eqnarray}\label{eq:11}
\frac{1}{\mu_{h(l)}}
(\vec{\nabla}\times\vec{B}_{h(l)})\times\vec{B}_{h(l)}=\frac{1}{\mu_{h(l)}}(\vec{B}_{h(l)}.\vec{\nabla})\vec
{B}_{h(l)}-\frac{1}{2\mu_{h(l)}} \vec{\nabla}(\vec{B^2}_{h(l)})
\end{eqnarray}

Substitution for $\vec{v}_{h(l)}$ from Eq. (8) in Eq. (9) followed
by use of Eq. (11) leads to Bernoulli's equation for the MHD fluid
\begin{eqnarray}\label{eq:12}
-\frac{\partial \phi_{h(l)}}{\partial t}+ \frac{1}{2}(\vec{\nabla}
\phi_{h(l)})^2=-\frac{p_{h(l)}}{\rho_{h(l)}}-gy-\frac{1}{2\mu_{h(l)}\rho_{h(l)}}B_{h(l)}^2+\frac{f_{h(l)}(t)}{\rho_{h(
l)}}
\end{eqnarray}

\section*{III. KINEMATICAL AND DYNAMICAL BOUNDARY CONDITIONS}

The kinematical boundary conditions satisfied by the interfacial
surface $y=\eta(x,t)$ are
\begin{eqnarray}\label{eq:13}
\frac{\partial \eta}{\partial t}+(v_h)_{x}\frac{\partial
\eta}{\partial x}=(v_h)_{y}
\end{eqnarray}
\begin{eqnarray}\label{eq:14}
(v_h)_{x}\frac{\partial \eta}{\partial x}-(v_l)_{x}\frac{\partial
\eta}{\partial x}=(v_h)_{y}-(v_l)_{y}
\end{eqnarray}

From Bernoulli's Eq. (12) for the heavier and lighter fluids one
obtains the following equation
\begin{eqnarray}\label{eq:15}
\nonumber\rho_h[-\frac{\partial \phi_h}{\partial t}+
\frac{1}{2}(\vec{\nabla} \phi_h)^2]-\rho_l[-\frac{\partial
\phi_l}{\partial t}+ \frac{1}{2}(\vec{\nabla}
\phi_l)^2]=-[g(\rho_h-\rho_l)y+(p_h-p_l)\\
+(\frac{{B^2}_{h}}{2\mu_{h}}-\frac{{B^2}_{l}}{2\mu_{l}})]+f_h(t)-f_l(t)
\end{eqnarray}

Further with the help of Eqs. (1) and (2) and the
incompressibility condition $\vec{\nabla}.\vec{v_{h(l)}}=0$, Eq.
(11) simplifies to
\begin{eqnarray}\label{eq:16}
\frac{\partial [\vec{B}_{h(l)}(x,y,t)]}{\partial
t}+(\vec{v}_{h(l)}.\vec{\nabla})\vec{B}_{h(l)}=0
\end{eqnarray}

The interfacial kinematic boundary conditions (13) and (14)
together with Bernoulli's Eq. (15) and magnetic induction Eq. (16)
are employed in the next section to obtain the temporal evolution
of the elevation of the tip of bubble (spike) like structures at
the two fluid interface from its undisturbed level.

\section*{IV. EQUATION FOR RAYLEIGH - TAYLOR INSTABILITY INDUCED INTERFACIAL STRUCTURE PARAMETERS}

Substituting $\eta(x,t)$ and $\phi_{h(l)}(x,y,t)$ from
Eqs.(3),(6)-(8) in Eqs. (13) and (14) and expanding in powers of
the transverse coordinate x up to i=2 and neglecting terms
O($x^{i}$)($i\geq3$), we obtain the following equations
$^{\cite{mg09}}$

\begin{eqnarray}\label{eq:17}
\frac{d\xi_1}{d\ t}=\xi_3
\end{eqnarray}
\begin{eqnarray}\label{eq:18}
\frac{d\xi_2}{d\ t}=-\frac{1}{2}(6\xi_2+1)\xi_3
\end{eqnarray}
\begin{eqnarray}\label{eq:19}
b_0=-\frac{6\xi_2}{(3\xi_2-\frac{1}{2})}ka_1
\end{eqnarray}
\begin{eqnarray}\label{eq:20}
b_1=\frac{(3\xi_2+\frac{1}{2})}{(3\xi_2-\frac{1}{2})}a_1
\end{eqnarray}
\begin{eqnarray}\label{eq:21}
\xi_1=k\eta_0; \qquad \xi_2=\eta_2/k; \qquad \xi_3=k^2a_1
\end{eqnarray}

$\xi_1$ and $\xi_2$ are respectively the nondimensionalized (with
respect to the wave length) displacement and curvature of the tip
of the bubble (spike) and $\xi_3/k$ is tip velocity.

At this stage it is in order to justify neglect of contribution
from terms of order $x^i (i\geq 3)$ as done here. This is provided
on two counts:

 (i) The interface displacement $y(x,t)$ is expanded
in Eq. (3) keeping only terms of order $x^2$,-the customary
practice in Layzer's approach. Since we are interested only in the
motion close to the tip of the bubble or spike,i.e., for $x\approx
0$ it is sufficient to retain terms up to order $x^2$ and neglect
$O(x^i)(i\geq 3)$.

(ii) Even if $\eta(x,t)$ is expanded as
\begin{eqnarray}
\nonumber\eta(x,t)=\eta_0(t)+\eta_2(t)x^2+\eta_4(t)x^4+\eta_6(t)x^6....
\end{eqnarray}
it can be shown that at the saturation level $(d\eta_i/dt=0)$
contributions from terms containing $\eta_4,\eta_6$...are much
smaller than that from $\eta_2(t)$(see Appendix). Thus expansion
of the kinematic condition and in its turn the expansion in
Bernoulli's equation and Faraday's equation (which follows later)
retaining higher order terms $O (x^4)$ can also be neglected.

 Next let us turn to the magnetic field induction
Eq. (16). To satisfy Eq. (16) with $\vec{v}_h$ given by Eq. (8) we
set
\begin{eqnarray}\label{eq:22}
B_h(x,y,t)=\beta_{h0}(t)+\beta_{h}(t)\cos{(kx)}e^{-k(y-\eta_0(t))};
\quad y>0
\end{eqnarray}
in Eq. (16); this leads to
\begin{eqnarray}\label{eq:23}
\dot{\beta}_{h0}(t)+(\dot{\beta}_{h}(t)+\beta_{h}(t)
k\dot{\eta}_0)\cos{(kx)}e^{-k(y-\eta_0(t))}-k^2a_1\beta_{h}e^{-2k(y-\eta_0(t))}=0
\end{eqnarray}

Corresponding to the parabolic interfacial structure represented
by $y(x,t)=\eta_0(t)+\eta_2(t)x^2$ the foregoing equation yields
on equating coefficients of $x^{i}$ $(i=0,2)$ and neglecting terms
O($x^i$)with $i\geq3$ the following relation

$i=0:$  \quad   $\dot{\beta}_{h0}(t)+\dot{\beta}_{h}(t)=0$

so that\begin{eqnarray}\label{eq:24}
\beta_{h0}(t)+\beta_{h}(t)=constant=B_{h0}, say
\end{eqnarray}

$i=2:$
\begin{eqnarray}\label{eq:25}
\frac{\delta\dot{B}_{h}}{\delta
B_{h}(t)}=\frac{(\xi_2-\frac{1}{2})}{(\xi_2+\frac{1}{2})}\xi_3;\quad
\delta B_{h}(t)=\frac{\beta_{h}(t)}{B_{h0}}
\end{eqnarray}
\begin{eqnarray}\label{eq:26}
\delta{B}_{h}(t)=\delta{B}_{h}(t=0)\exp\left[{\int
_0^t\xi_3\frac{(\xi_2-\frac{1}{2})}{(\xi_2+\frac{1}{2})}}d\tau\right]
\end{eqnarray}
so that $\delta B_{h}(t=0)>(<0)$; according as $\delta
B_{h}(t=0)>(<0)$.

In obtaining Eqs. (24) and (25) we have used the relation
$\xi_3=\dot{\xi_1}=k\dot{\eta_0}=k^2a_1$\quad(Eq. (17)).

Similarly, to satisfy the magnetic field induction equation in the
region $y<0$, i.e., in the region occupied by the lighter fluid we
set
\begin{eqnarray}\label{eq:27}
B_l(x,y,t)=\beta_{l0}(t)+\beta_{l}(t)\cos{(kx)}e^{k(y-\eta_0(t))};
\end{eqnarray}
and proceeding as in case of the magnetic field induction
$B_h(x,y,t)$in region $y>0$, we obtain
\begin{eqnarray}\label{eq:28}
\beta_{l0}(t)+\beta_{l}(t)=constant=B_{l0}, say
\end{eqnarray}
and
\begin{eqnarray}\label{eq:29}
\frac{\delta\dot{B}_{l}}{\delta
B_{l}(t)}=\frac{(\xi_2+\frac{1}{2})}{(\xi_2-\frac{1}{2})}\frac{(\xi_2+\frac{1}{6})}{(\xi_2-\frac{1}{6})}\xi_3;\quad
\delta B_{l}(t)=\frac{\beta_{l}(t)}{B_{l0}}
\end{eqnarray}
by using Eqs. (19) and (20)($\Longrightarrow b_0+kb_1+ka_1=0$).

Again proceeding as in the deduction of Eq. (26) we obtain
\begin{eqnarray}\label{eq:30}
\delta B_{l}(t)=\delta
B_{l}(t=0)exp\left[{\int_0^t\xi_3\frac{(\xi_2+\frac{1}{2})}{(\xi_2-\frac{1}{2})}}\frac{(\xi_2+\frac{1}{6})}{(\xi_2-\frac{1}{6})}d\tau\right]
\end{eqnarray}
so that $ \delta B_{l}(t=0)>(<0)$; according as $ \delta
B_{l}(t=0)>(<0)$.

The magnetic field affected Rayleigh - Taylor instability induced
growth of the mushroom shaped surface structure are determined by
the parameters $\xi_1(t),\xi_2(t),\xi_3(t)$ as also the magnetic
induction perturbation $\delta B_h(t)$ and $\delta B_l(t)$. To
determine the time evolution of these five functions we need aside
from the differential Eqs. (17),(18),(25) and (29) an extra one to
complete the set. This is provided by Eq. (15). Now using Eqs.
(22) and (24) one obtains
\begin{eqnarray}\label{eq:31}
\frac{1}{2\mu_{h}}B^2_{h}(x,y,t)=\frac{B^2_{h0}}{2\mu_{h}}-k^2\frac{{B^2}_{h0}}{\mu_{h}}\delta
B_{h}(t)(\xi_2+\frac{1}{2})x^2
\end{eqnarray}
Similarly using Eqs. (27) and (28) one obtains
\begin{eqnarray}\label{eq:32}
\frac{1}{2\mu_{l}}B^2_{l}(x,y,t)=\frac{B^2_{l0}}{2\mu_{l}}+k^2\frac{{B^2}_{l0}}{\mu_{l}}\delta
B_{l}(t)(\xi_2-\frac{1}{2})x^2
\end{eqnarray}
where
\begin{eqnarray}\label{eq:33}
\mid \delta B_{h}(t)\mid , \mid
\delta B_{l}(t)\mid\ll 1
\end{eqnarray}
whenever the initial values $\mid \delta B_h(0)\mid$ and
$\mid\delta B_l(0)\mid \ll 1 $ as may be seen from Eqs. (26) and
(30). This anticipation is substantiated later by numerical
computation (Fig. 2 -Fig. 5).

We next substitute for
$B^2_{h}(x,y,t)/2\mu_h-B^2_{l}(x,y,t)/{2\mu_{l}}$ from Eqs. (31)
and (32) in Eq. (16) and use the dynamical boundary condition
expressing balance of fluid and finite order magnetic pressure on
two sides of the interface:
\begin{eqnarray}\label{eq:34}
p_h+\frac{B^2_{h0}}{2\mu_h}=p_l+\frac{B^2_{l0}}{2\mu_l}
\end{eqnarray}
Eq. (15) now reduces to
\begin{eqnarray}\label{eq:35}
\nonumber\rho_h[-\frac{\partial \phi_h}{\partial t}+
\frac{1}{2}(\vec{\nabla} \phi_h)^2]-\rho_l[-\frac{\partial
\phi_l}{\partial t}+\frac{1}{2}(\vec{\nabla}
\phi_l)^2]=-g(\rho_h-\rho_l)y+k^2\frac{B^2_{h0}}{\mu_h}\delta
B_h(t)(\xi_2+\frac{1}{2})x^2\\
+\frac{B^2_{l0}}{\mu_l}\delta
B_l(t)(\xi_2-\frac{1}{2})x^2+f_h(t)-f_l(t)
\end{eqnarray}
which involves the influence only of the infinitesimal magnetic
field fluctuation on the interfacial structure. After some lengthy
but straightforward algebraic manipulation we arrive at the
required equation which is the last Eq. of the following set of
Eqs. (36). The last Eq. of the set of Eqs. (36) is the required
one as mentioned before and represents the dynamical boundary
condition and obtained by setting $y=\eta_0+\eta_2x^2$ and
equating coefficient of $x^2$ on both sides. All the equations are
collected together below for the sake of convenience.

\parbox{11cm}{\begin{eqnarray*}
\frac{d\xi_1}{d \tau}=\xi_3/\sqrt{kg} \hskip 750pt\\
\frac{d\xi_2}{d \tau}=-\frac{1}{2}(6\xi_2+1)\xi_3/\sqrt{kg}  \hskip 690pt \\
\frac {\frac{d }{d \tau }{\delta B_h(t)}}{{\delta
B_h(t)}}=\frac{(\xi_2-\frac{1}{2})}{(\xi_2+\frac{1}{2})}\xi_3/\sqrt{kg}
\hskip
685pt \\
\frac {\frac{d }{d \tau }{\delta B_l(t)}}{{\delta
B_l(t)}}=\frac{(\xi_2+\frac{1}{2})}{(\xi_2-\frac{1}{2})}
\frac{(\xi_2+\frac{1}{6})}{(\xi_2-\frac{1}{6})}\xi_3/\sqrt{kg} \hskip 645pt \\
\frac{d\xi_3}{d\tau}=-\frac{N(\xi_2,r)}{D(\xi_2,r)}\frac{(\xi_3/\sqrt{kg})^2}{(6\xi_2-1)}+2(r-1)\frac{\xi_2(6\xi_2-1)}{D(\xi_2,r)} \hskip 575pt\\
-\frac{(6\xi_2-1)}{D(\xi_2,r)}[r\frac{k V^2_h}{g}\delta
B_h(t)(2\xi_2+1)+\frac{k V^2_l}{g}\delta B_l(t)(2 \xi _2-1)]
\hskip 500pt
\end{eqnarray*}}\hfill
\parbox{0.1cm}{\begin{eqnarray}\label{eq:36} \end{eqnarray}} \\
\begin{eqnarray}\label{eq:37}
where,\nonumber  \tau=t\sqrt{kg}; \quad r=\frac{\rho
_h}{\rho_l};\quad
D(\xi_2,r)=12(1-r)\xi_{2}^{2}+4(1-r)\xi_{2}+(r+1); \\
N(\xi_2,r)=36(1-r)\xi_{2}^{2}+12(4+r)\xi_{2}+(7-r)
\end{eqnarray}
\begin{eqnarray}\label{eq:38}
V_{h(l)}=\sqrt{B^2_{h0(l0)}/\rho_{h(l)}\mu _{h(l)}}
\end{eqnarray}
is the Alfven velocity in the heavier (lighter) fluid.

The above set of Eqs. describe the time evolution of a bubble. The
time evolution of a spike is obtained from the same set by making
the transformation $\xi_1\rightarrow-\xi_1,\xi_2\rightarrow-\xi_2$
and $r\rightarrow\frac {1}{r}$$^{\cite{vg02}}$. It is important to
note that in the last Eq. of the set of Eqs.(36) the contribution
to the bubble tip velocity $\frac{d\xi_3}{d\tau}$ from the force
of buoyancy $g(\rho_h-\rho_l)$y  is proportional to
$kg(\rho_h-\rho_l)\xi_2$ while that from the magnetic pressure
fluctuation are proportional to $k^2(B^2_{l0}/\mu_{l})\delta
B_{l}(t)(\xi_2-\frac{1}{2})$ and $k^2(B^2_{h0}/\mu_{h})\delta
B_{h}(t) (\xi_2+\frac{1}{2})$ respectively as may be seen from
Eqs. (31) and(32). Further both for bubbles and spikes $\xi_2$
lies in $(-\frac{1}{6},\frac{1}{6})$; hence we always have
$(\xi_2-\frac{1}{2})< 0 $ and $(\xi_2+\frac{1}{2})> 0$. So by
applying condition Eq. (30)  we find
\begin{eqnarray}\label{eq:39}
k^2\frac{B^2_{l0}}{\mu_l}\delta B_{l}(t)(\xi_2-\frac{1}{2})< or
> 0 ;
\end{eqnarray}
according as $\delta B_{l}(t=0)> or < 0$.

Similarly on applying Eq. (26) it follows that
\begin{eqnarray}\label{eq:40}
k^2\frac{B^2_{h0}}{\mu_h}\delta B_{h}(t)(\xi_2+\frac{1}{2})> or <
0;
\end{eqnarray}
according as $\delta B_{h}(t=0)> or < 0$.

\section*{V. RESULTS AND DISCUSSIONS}

Analytical closed form solution of the set of Eqs. (36) not being
feasible we take recourse to the method of numerical solution (5th
order Runge-Kutta-Fehlberg method) and consider the following
cases.

\textbf{\underline{Case A}}

Assume $B_{h0}=0$, $B_{l0}\neq 0$. Such a situation may occur when
the lighter fluid (occupying the lower region $y< 0$) is ionized
while the heavier fluid (region $y > 0 $) is nonmagnetic. From
Eqs. (32) and (36) and the concluding discussions of the foregoing
section it is clearly seen that the instability driving pressure
difference $g(\rho_h -\rho_l)\xi_2$ is lowered or enhanced by
$\mid k^2\frac{B^2_{l0}}{\mu_l}\delta
B_{l}(t)(\xi_2-\frac{1}{2})|$ according as $\delta B_{l}(t=0)$ is
$ > 0$ or $ < 0 $. The concomitant growth rate modifications are
shown in Fig. 2 (Fig. 3) which plots the bubble (spike) tip
elevation $\mid \xi_1\mid$ and growth rate $\mid \dot{\xi_1}\mid
$. Fig. 2 and Fig. 3 show that whether in case of suppression or
enhancement the growth rate  $\xi_3(=\mid\dot{\xi_1}\mid)$
approaches an asymptotic value as $\tau \rightarrow \infty$ both
for bubble and for spike. This happens as $\delta B_l(t)$ exhibits
similar asymptotic behavior as one may see in Fig. 2 and Fig. 3.
The following analytic expressions for $(\xi_3)_{asymp}$ as $\tau
\rightarrow \infty$ are obtained by setting $d\xi_3/d\tau=0$
together with $B_{h0}=0$ in the last Eq. of the set of Eqs. (36):
\begin{eqnarray}\label{eq:41}
[(\xi_3)_{asymp}]_{bubble}=\sqrt{\frac{2Akg}{3(1+A)}}\sqrt{1-2(\frac{1-A}{A})\frac{kV^2_l}{g}[\delta
B_l(\infty)]_{bubble}}
\end{eqnarray}
\begin{eqnarray}\label{eq:42}
[(\xi_3)_{asymp}]_{spike}=\sqrt{\frac{2Akg}{3(1-A)}}\sqrt{1-2(\frac{1+A}{A})\frac{kV^2_l}{g}[\delta
B_l(\infty)]_{spike}}
\end{eqnarray}

Here $\delta B_l(\infty)$denotes the asymptotic value. The growth
rate increases (destabilization) if $\delta B_l(0)<0$ (hence
$\delta B_l(\infty)<0)$,i.e., the magnetic field perturbation
diminishes the pressure below the interface relative to that
above. On the other hand $(\xi_3)_{asym}$ decrease and asymptote
to $0$ (zero) as $kV^2_l/g$ increases if $\delta B_l(0)>0$ and
therefore $\delta B_l(\infty)>0$,i.e., the pressure below the
interface increases and tends to restore stability.

\textbf{\underline{Case B}}

Assume $B_{h0}\neq 0$ but $B_{l0}= 0$. This situation is the
reverse of that in case \textbf{A} and may arise when the heavier
fluid is ionized while the lighter one is non magnetic. The
dynamical boundary condition shows that following the same line of
arguments as in case \textbf{A} but with $B_{h0}\neq 0$ but
$B_{l0}= 0$ we find that the instability driving force $g(\rho_h
-\rho_l)\xi_2$ is now enhanced or reduced by $\mid
k^2\frac{B^2_{h0}}{\mu_h}\delta B_{h}(t)(\xi_2+\frac{1}{2})| $
according as $\delta B_{h}(t=0)$ is $ > $ or $ < $ 0. This
conclusion is supported by the difference in the height of the
bubble (or spike) tip shown in Fig. 4 and Fig. 5. However note
that $\delta B_h(t)\rightarrow 0$ as $t\rightarrow\infty$. This
has the consequence that the asymptotic value of the velocity of
the tip of the bubble (in spike) height $(\xi_3)_{asymp}$ is the
same as in the absence of magnetic field. But as
$\mid\dot{\xi_1}\mid=\xi_3$ the height of the tip of the bubble
(or spike) maintains a constant difference.

\textbf{\underline{Case C}}
 Assume both fluids are conducting and
magnetic field is non zero on either side. We have considered two
cases

(i) $r\frac{k}{g}V^2_{h}=\frac {k}{g}V^2_l=5.0$

(ii)$r\frac{k}{g}V^2_{h}=5.0,\frac {k}{g}V^2_l=10.0$

(iii)$r\frac{k}{g}V^2_{h}=\frac {k}{g}V^2_l=1.2$

(iv)$r\frac{k}{g}V^2_{h}=\frac {k}{g}V^2_l=1.4$

with $\delta B_{h}(t=0)=\delta B_{l}(t=0) > 0$ in each case. The
bubble tip elevation $\xi_1$ as well as its velocity
$\dot{\xi_1}=\xi_3$ oscillates as the magnetic pressure acts on
both sides of the interfaces but in opposite direction and with
opposite phase. The results are shown in Fig. 6. In (i) and also
in (iii) and (iv) the growth rate $\xi_3=\dot{\xi_1}$ oscillates
approximately symmetrically about  $\xi_3=0$ as
$r\frac{k}{g}V^2_{h}=\frac {k}{g}V^2_l$ equation for in
$\dot{\xi_3}$ in set of Eqs. (36); in (ii) the asymmetry results
from difference in the driving pressure difference on two sides.
Moreover it is to be noted from Fig. 7 and Fig. 8 that the
oscillation frequency increases with 'k' and also with Alfven
velocity. Occurrence of such an oscillation were also concluded
for RMI $^{\cite{zq08} ,\cite{jc09}}$ with increase in frequency
similar to our case; however such oscillation are harmonic as
against the nonlinear oscillations in our case.

\section*{VI. SUMMARY}
Finally we summarize the results :

The change in interfacial pressure difference due to magnetic
field fluctuation leads to enhancement or suppression of
instability as stated below.

(i) If $B_{h0}=0$,i.e, there exists no magnetic field above the
two fluid interface $(y>0)$ but $B_{l0}\neq 0$ the lowering of the
magnetic field below the interface $y=0$ due to an initial
perturbation $\delta B_l(0)<0 (\Rightarrow \delta B_l(t)<0$ (by
Eq. (30)) according to the fourth Eq. of the set of Eqs. (36))
leads to depression of pressure on the side of the lighter fluid
with the result that the instability growth is enhanced (Fig. 2).
On the other hand if the initial perturbation $\delta B_l(0)>0
(\Rightarrow\delta B_l(t)>0$ (by Eq. (30)) the pressure on the
side of  the lighter fluid increases with resulting suppression of
growth rate which asymptote to $0$ (zero) as $(\frac{kV^2_l}{g})$
increases (Fig. 3).

$\delta B_l(t)$ tends to a constant value asymptotically as
$t\rightarrow \infty$; this enables us to obtain an analytic
expression for the asymptotic growth rate $(\xi_3)_{asym}$ both
for bubble and spike as given by Eqs. (41) and (42) respectively.

(ii) If $B_{l0}=0$ but $B_{h0}\neq 0$, it is $\delta
B_h(t)\rightarrow 0$ (asymptotically whether  initial perturbation
$\delta B_h(0)>0$ or $<0$ (Fig. 4 and Fig. 5). This has the
consequence that the asymptotic growth rate becomes the same as in
the nonmagnetic case.

(iii) If both $B_{h0}\neq 0$and $B_{l0}\neq 0$ with $\delta
B_h(t=0)=\delta B_l(t=0)>0$ the magnetic pressure perturbation
acts on both sides of the interface but in opposite direction with
opposite phase. This has the consequence that the growth rate
$\xi_3=\dot{\xi_1}$ oscillates symmetrically about $\xi_3=0$ if
$\frac{kV^2_h}{g}=\frac{kV^2_l}{g}$ which increases in amplitude
and frequency as the Alfven velocity increases. However, if
$\frac{kV^2_h}{g}\neq \frac{kV^2_l}{g}$ the oscillation is
asymmetrical about $\xi_3=0$.

\section*{ACKNOWLEDGEMENTS}
This work is supported by the Department of Science \& Technology,
Govt. of India under grant no. SR/S2/HEP-007/2008. The authors are
thankful to the referee for his helpful critical comments which
provides improvement in the presentation of the paper.

\section*{Appendix:}

Let the surface displacement $\eta(x,t)$ in Layzer's model
expanded retaining higher powers of x:-
$\eta(x,t)=\eta_0(t)+\eta_2(t)x^2+\eta_4(t)x^4+\eta_6(t)x^6.....$

The time dependence of the coefficient functions $\eta_i(t)$ one
obtained by equating coefficient of $x^i (i=0,2,4,6...)$ in the
expansion of the kinematical condition in powers of x:-

$\frac{\partial \eta(x,t)}{\partial t}-\frac{\partial
\phi}{\partial x}\frac{\partial \eta(x,t)}{\partial
x}=-\frac{\partial \phi}{\partial y}$

where the velocity potential
$\phi(x,y,t)=a_1(t)\cos{(kx)}e^{-k(y-\eta_0(t))}$

This gives coefficient  of $x^2$:

$\frac{d\eta_2}{dt}=-ka_1\left[3k\eta_2+\frac{k^2}{2!}\right]$

Coefficient of $x^4$:
$\frac{d\eta_4}{dt}=-ka_1\left[5k\eta_4-\frac{5}{2}k^2\eta^2_2
-\frac{5}{6}k^3\eta_2 -\frac{k^4}{4!}\right]$

Coefficient of $x^6$:
$\frac{d\eta_6}{dt}=-ka_1\left[7k\eta_6-7k^2\eta_2\eta_4+\frac{7k^4\eta^2_2}{12}+\frac{7k^5\eta_2}{120}-\frac{7k^3\eta_4}{6}+\frac{7k^3\eta^3_2}{6}+\frac{k^6}{6!}\right]$

giving saturation values $(d\eta_i/dt=0)$:

$k\eta_2=-k^2/6$;

$k\eta_4=-k^4/180$;

$k\eta_6=-k^6/2835$....

\begin{figure}[p]
\vbox{\hskip 1.5cm \epsfxsize=12cm \epsfbox{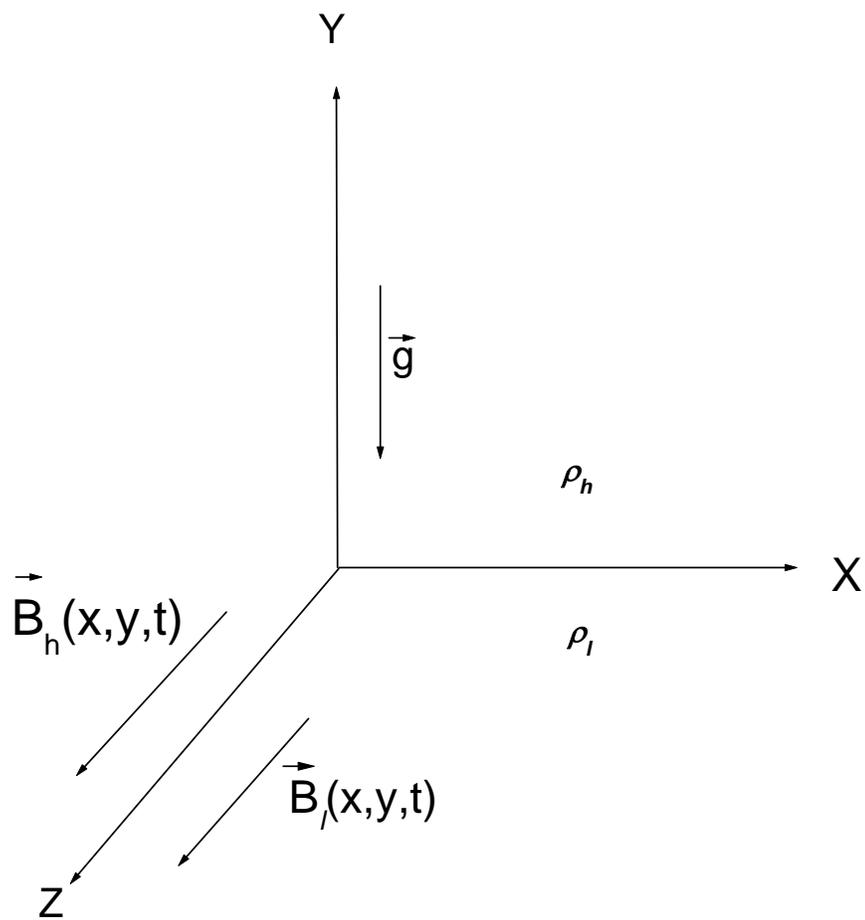}}
\begin{verse}
\vspace{-0.25cm} \caption{Geometry of the model} \label{fig:1}
\end{verse}
\end{figure}

\begin{figure}[p]
\vbox{ \hskip 1.5cm \epsfxsize=12cm \epsfbox{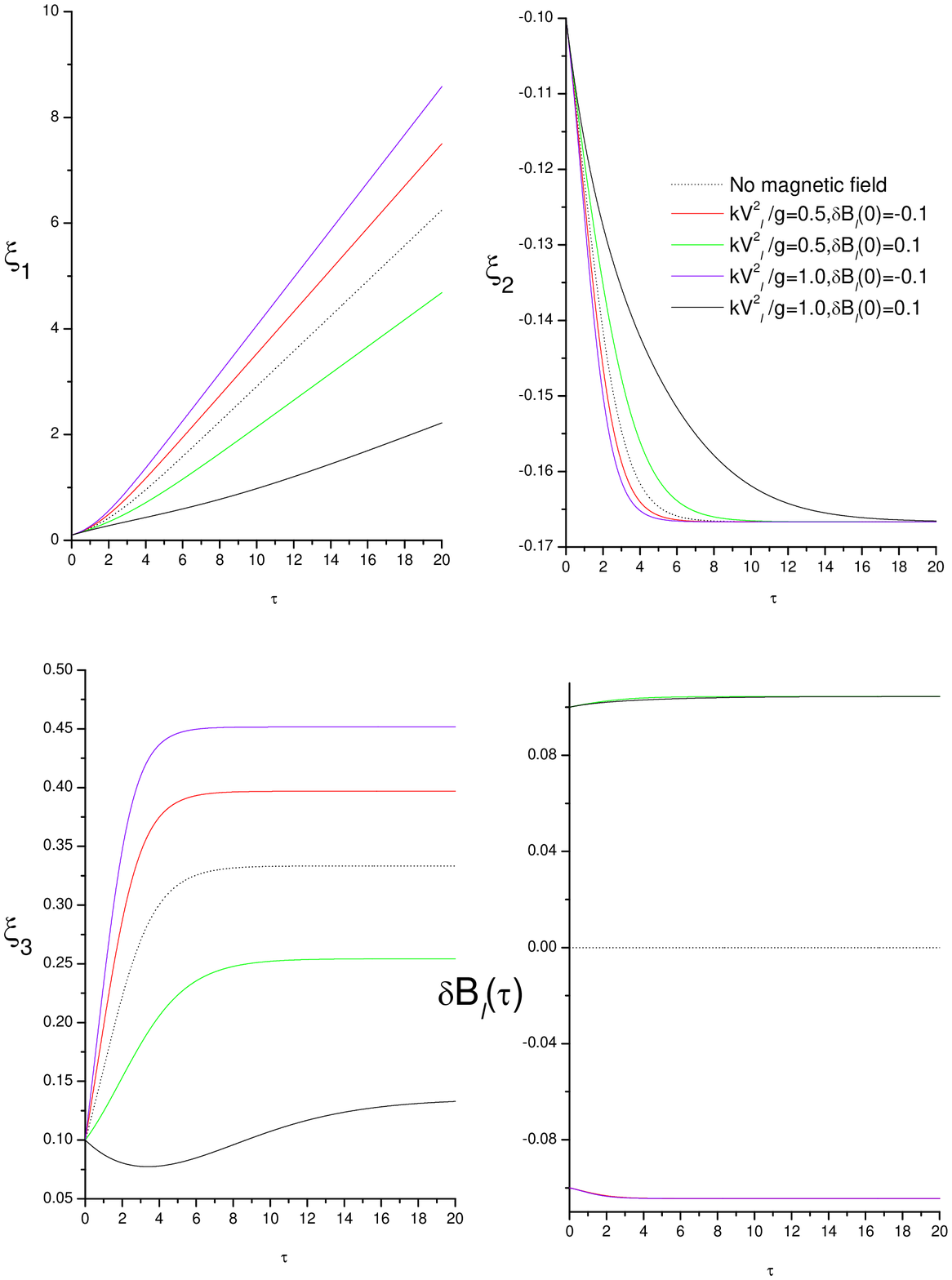}}
\begin{verse}
\vspace{-0.25cm} \caption{Variation of $\xi_1$, $\xi_2$,bubble
growth rate  $\xi_3(=\dot\xi_1)$ and $\delta B_{l}$ with $\tau$
for $V^2_h =0$ [Eq. \ref{eq:36}]. Initial values
$\xi_1=0.1,\xi_2=-0.1,\xi_3=0.1$ and $r=1.5$ } \label{fig:2}
\end{verse}
\end{figure}

\begin{figure}[p]
\vbox{ \hskip 1.5cm \epsfxsize=12cm \epsfbox{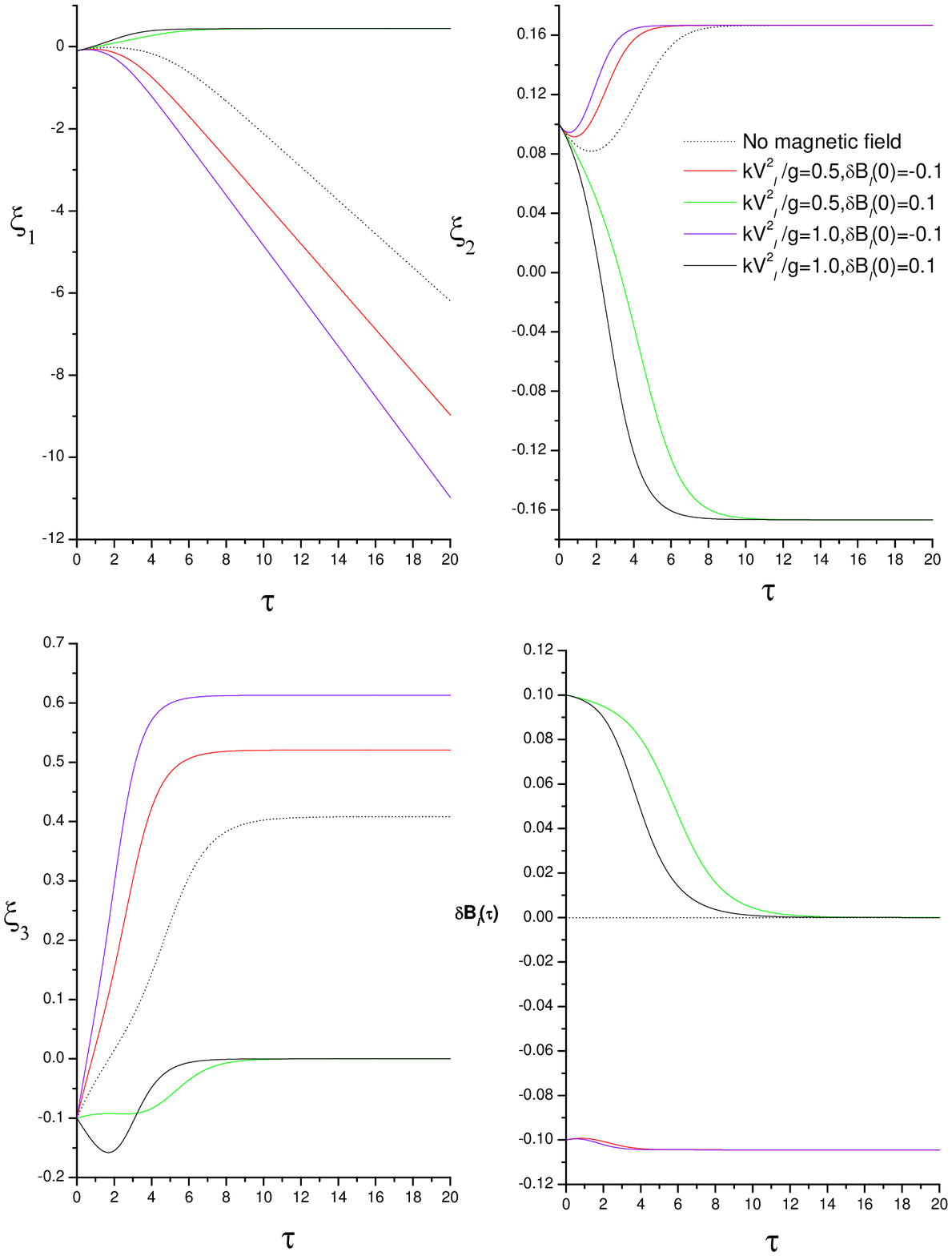}}
\begin{verse}
\vspace{-0.25cm} \caption{Variation of $\xi_1$, $\xi_2$, spike
growth rate $\xi_3(=\dot\xi_1)$ and $\delta B_{l}$ with $\tau$ for
$V^2_h =0$[Eq.36] (with transformation
$\xi_1\rightarrow-\xi_1,\xi_2\rightarrow-\xi_2,r\rightarrow 1/r$
in Eq.\ref{eq:36}).Initial values
$\xi_1=-0.1,\xi_2=0.1,\xi_3=-0.1$ and $r=1.5$ } \label{fig:3}
\end{verse}
\end{figure}

\begin{figure}[p]
\vbox{ \hskip 1.5cm \epsfxsize=12cm \epsfbox{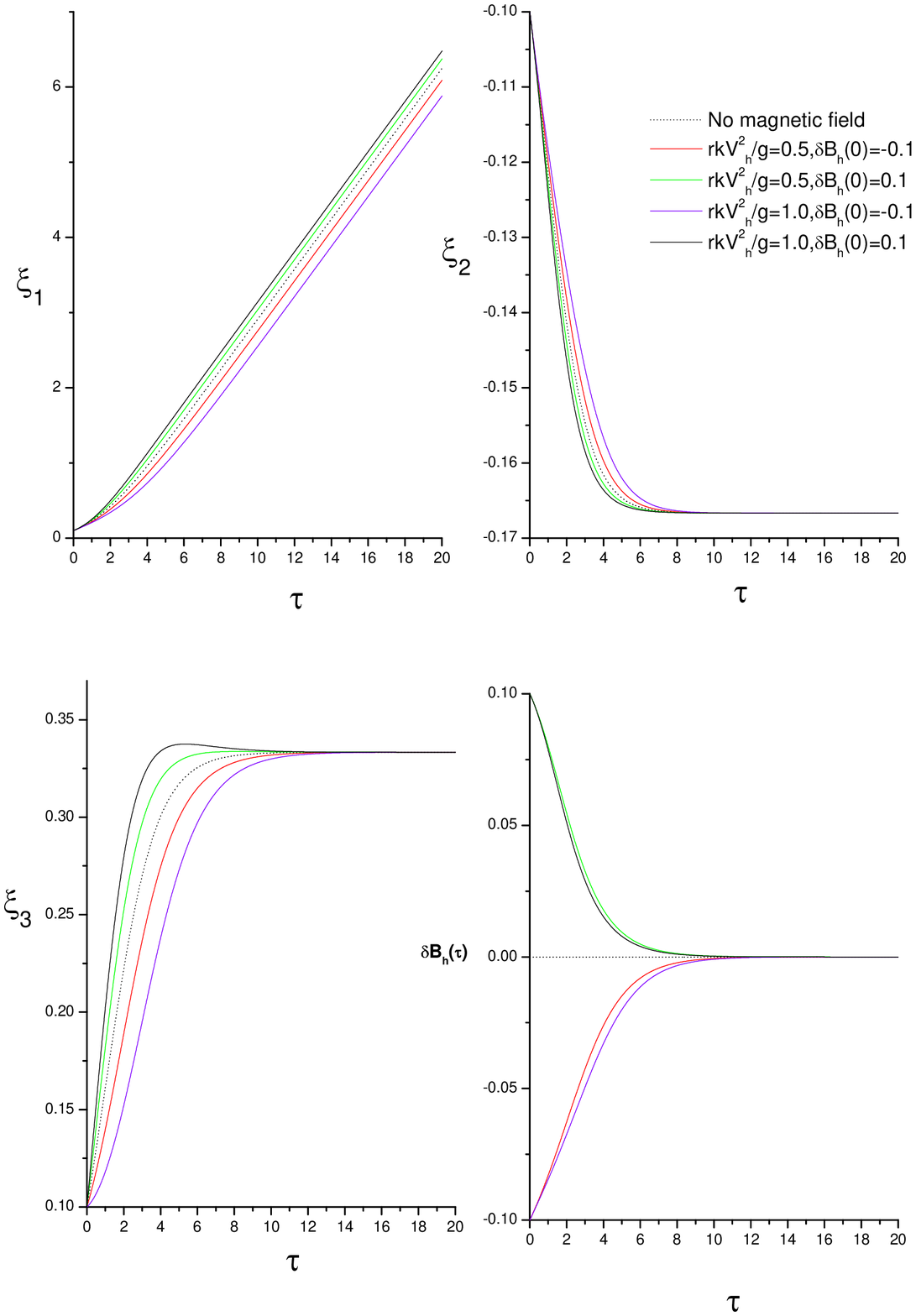}}
\begin{verse}
\vspace{-0.25cm} \caption{Variation of $\xi_1$, $\xi_2$, bubble
growth rate $\xi_3(=\dot\xi_1)$ and $\delta B_{h}$ with $\tau$ for
$V^2_l =0$[Eq.36 ].Initial values
$\xi_1=0.1,\xi_2=-0.1,\xi_3=0.1$, and $r=1.5$ } \label{fig:4}
\end{verse}
\end{figure}

\begin{figure}[p]
\vbox{ \hskip 1.5cm \epsfxsize=12cm \epsfbox{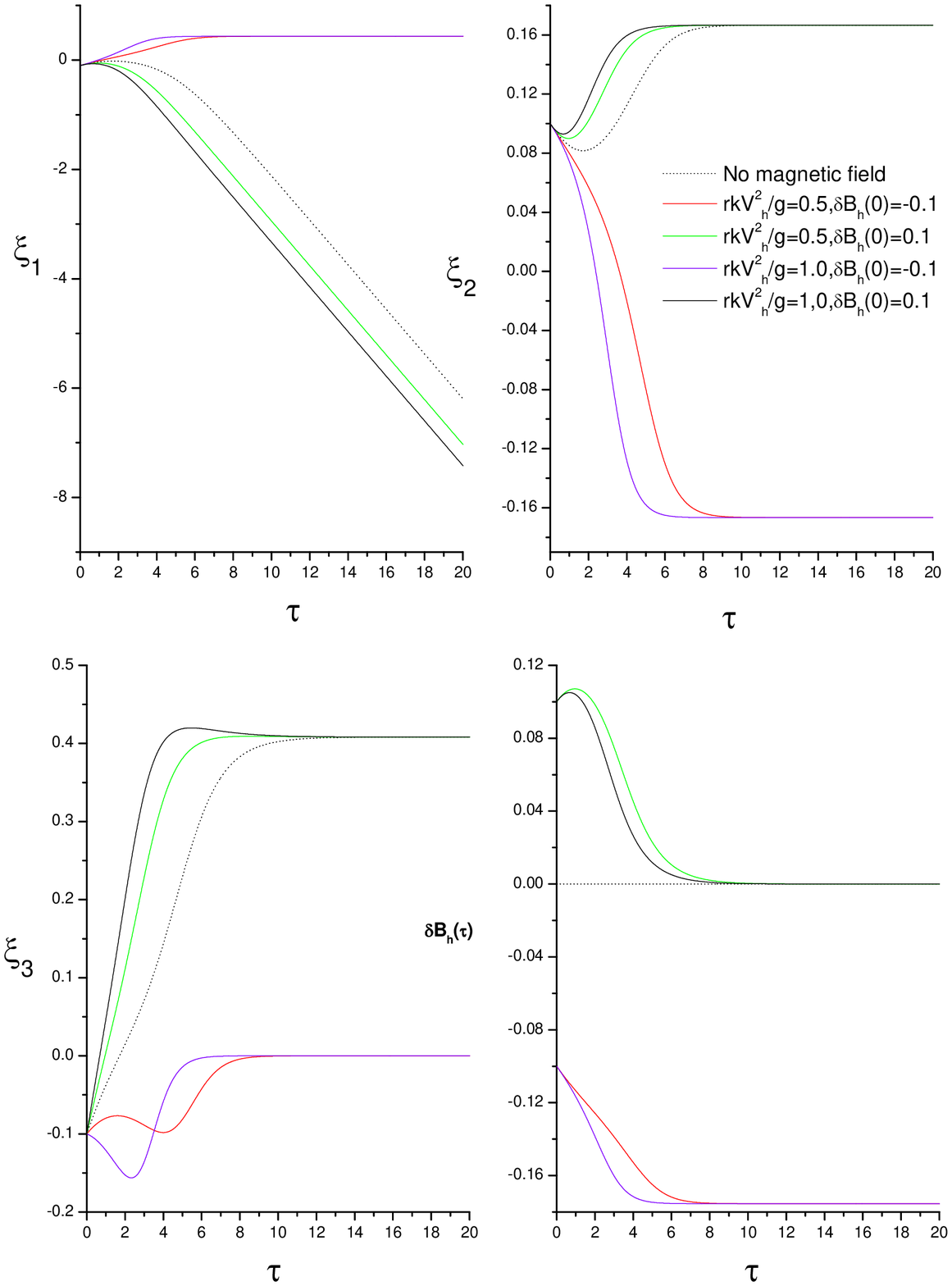}}
\begin{verse}
\vspace{-0.25cm} \caption{Variation of $\xi_1$, $\xi_2$, spike
growth rate $\xi_3(=\dot\xi_1)$ and $\delta B_{h}$ with $\tau$ for
$V^2_l =0$Eq.36 (with transformation
$\xi_1\rightarrow-\xi_1,\xi_2\rightarrow-\xi_2,r\rightarrow 1/r$
in in Eq.\ref{eq:36}).Initial values
$\xi_1=-0.1,\xi_2=0.1,\xi_3=-0.1$ and $r=1.5$ } \label{fig:5}
\end{verse}
\end{figure}

\begin{figure}[p]
\vbox{ \hskip 1.5cm \epsfxsize=12cm \epsfbox{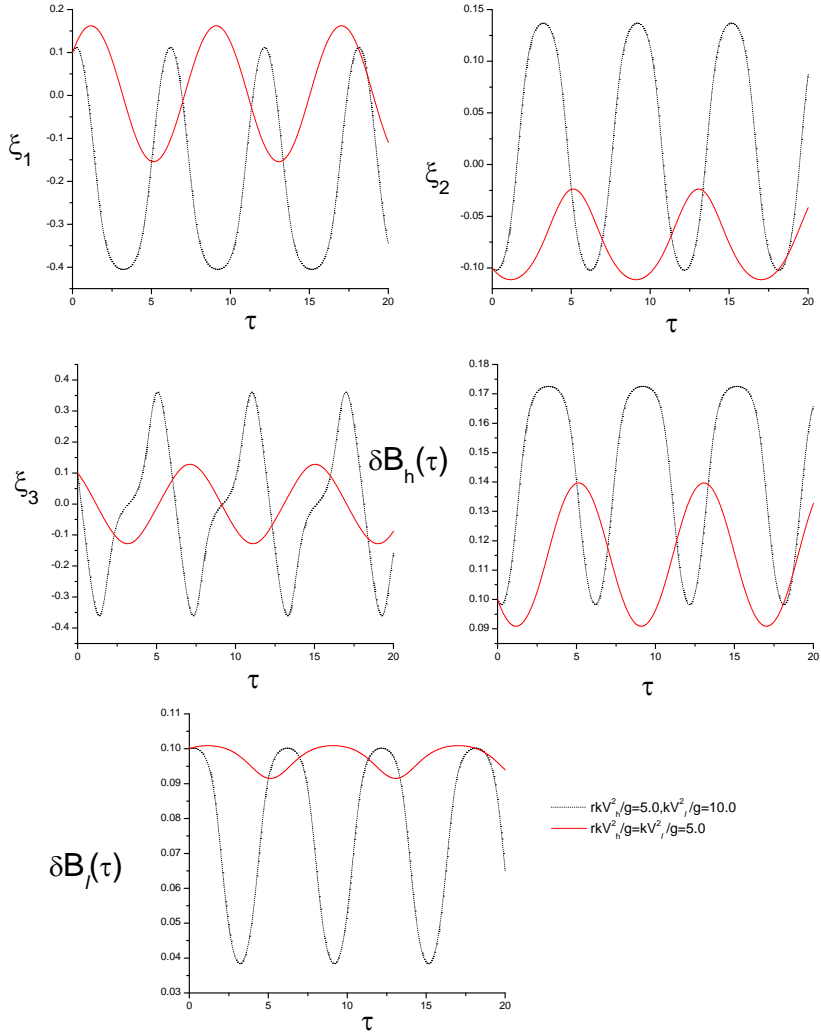}}
\begin{verse}
\vspace{-0.25cm} \caption{Growth rate oscillations for bubble.
Initial values $\xi_1=0.1,\xi_2=-0.1,\xi_3=0.1$,$\delta
B_{l}(0)=\delta B_{h}(0)=0.1$ and $r=1.5$ } \label{fig:6}
\end{verse}
\end{figure}

\begin{figure}[p]
\vbox{ \hskip 1.5cm \epsfxsize=12cm \epsfbox{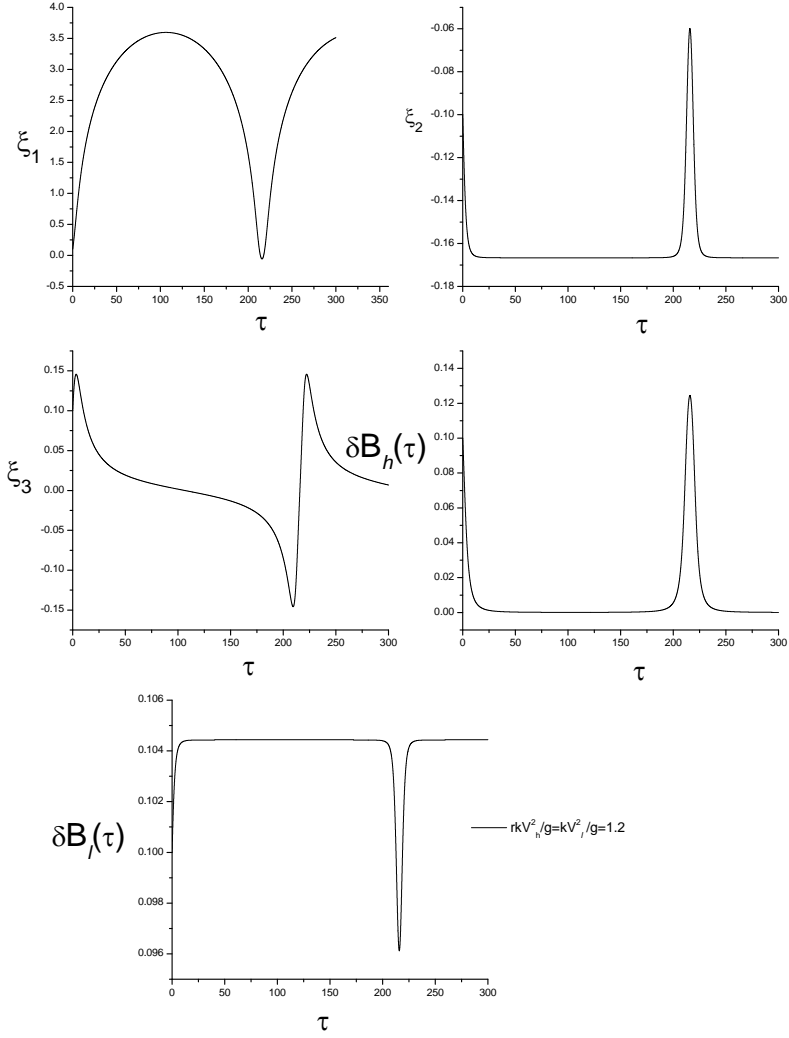}}
\begin{verse}
\vspace{-0.25cm} \caption{Oscillation of $\xi_1$, $\xi_2$, bubble
growth rate $\xi_3(=\dot\xi_1)$, $\delta B_{h}$ and $\delta B_{l}$
with $\tau$ as obtained by the solution of Eq. (\ref{eq:36}) with
initial values $\xi_1=0.1,\xi_2=-0.1,\xi_3=0.1$,$\delta
B_{l}(0)=\delta B_{h}(0)=0.1$ and $r=1.5$ } \label{fig:7}
\end{verse}
\end{figure}

\begin{figure}[p]
\vbox{ \hskip 1.5cm \epsfxsize=12cm \epsfbox{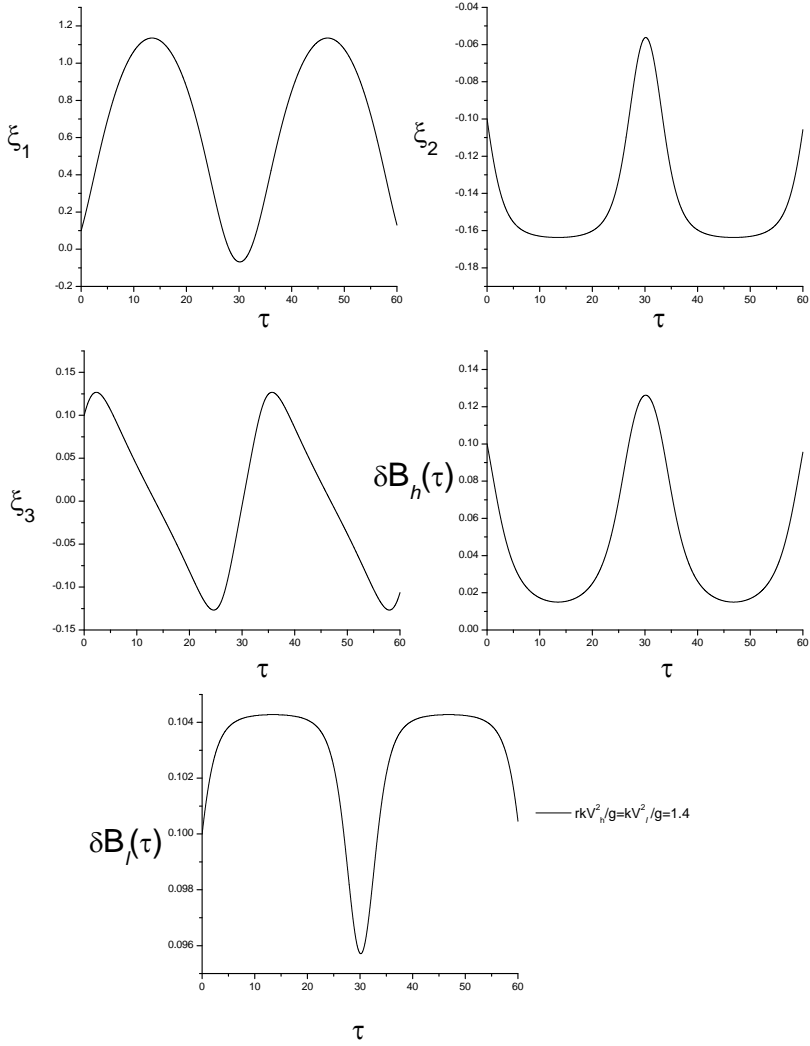}}
\begin{verse}
\vspace{-0.25cm} \caption{Growth rate oscillation for bubble as
obtained by the solution of Eq.(\ref{eq:36}) with initial values
$\xi_1=0.1,\xi_2=-0.1,\xi_3=0.1$,$\delta B_{l1}(0)=\delta
B_{h1}(0)=0.1$ and $r=1.5$ } \label{fig:8}
\end{verse}
\end{figure}

\end{document}